\documentclass{article}

\usepackage{graphicx} 
\usepackage{amsthm}

\usepackage{amsmath}
\usepackage{amsfonts}
\usepackage{amsthm}
\usepackage{mathtools}
\usepackage{siunitx}
\usepackage{float}

\usepackage{bm}

\usepackage{booktabs}
\usepackage{multirow}
\usepackage{ragged2e}
\usepackage{comment}
\usepackage{soul}
\usepackage{color}
\usepackage[normalem]{ulem}

\usepackage[bookmarks=true,colorlinks=true,linkcolor=red,anchorcolor=blue,citecolor=blue,urlcolor=blue]{hyperref}
\usepackage{listings}
\usepackage{natbib}
\usepackage[left=1.5in,right=1.5in,top=1in,bottom=1in]{geometry}
\def\E{{\mathbb E}}

\def\T{{ \mathrm{\scriptscriptstyle T} }}

\newtheorem{theorem}{Theorem}
\newtheorem{Assumption}[theorem]{\sc Assumption}

\begin{document}

\title{Robust inference for the unification of confidence intervals in meta-analysis}

\author{Wei Liang, Haicheng Huang,
\textit{
Xiamen University, China,}\\[2pt]
Hongsheng Dai$^\ast$, 
\textit{Newcastle University, Newcastle upon Tyne, U.K.}\\[2pt]
{hongsheng.dai@newcastle.ac.uk,}\\
and Yinghui Wei,
\textit{
University of Plymouth, U.K.}\\[2pt]
}

\markboth%
{Wei Liang and others}
{Robust inference for the unification of confidence intervals in meta-analysis}

\maketitle

\footnotetext{To whom correspondence should be addressed.}

\begin{abstract}
{
Traditional meta-analysis assumes that the effect sizes estimated in individual studies follow a Gaussian distribution. However, this distributional assumption is not always satisfied in practice,  leading to potentially biased results. In the situation when the number of studies, denoted as $K$, is large, the cumulative Gaussian approximation errors from each study could make the final estimation unreliable. In the situation when $K$ is small, it is not realistic to assume the random-effect follows Gaussian distribution.
In this paper, we present a novel empirical likelihood method for combining confidence intervals under the meta-analysis framework. This method is free of the Gaussian assumption in effect size estimates from individual studies and from the random-effects. We establish the large-sample properties of the non-parametric estimator, and introduce a criterion governing the relationship between the number of studies, $K$, and the sample size of each study, $n_i$. Our methodology supersedes conventional meta-analysis techniques in both theoretical robustness and computational efficiency. We assess the performance of our proposed methods using simulation studies, and apply our proposed methods to two examples. 
}
{confidence Interval; Empirical Likelihood; Meta-analysis; Random-effect Model.}
\end{abstract}

\section{Introduction}\label{sec:intro}

Meta-analysis is a statistical method to combine results from multiple studies, facilitating the inference-making process concerning specific parameters of interests, such as odds ratio. Parameters estimated from individual studies can be combined using either a fixed-effect  or a random-effects meta-analysis model \citep{HigginsandThompson2009, Deeks2022}.  A fixed-effect model assumes that a single parameter value is common to all studies, and a random-effects model assumes that parameters of the underlying studies follow some distributions. 

Let the random variable $Y_i$ $(i=1, 2, \cdots, K)$ denote an effect size estimate from study $i$.  Here $Y_i$ could be the mean difference between treatment groups, log-odds ratios, log-hazard ratios, or other effect size estimates from study $i$. The conventional random-effects meta-analysis model is defined as
\begin{equation}\label{eq:rma}
Y_i = \theta_i+ \epsilon_i, \;\;\;\;\;
\theta_i = \theta + \xi_i
\end{equation}
where $\theta_i$ is the true effect size of study $i$, $\theta$ is the common overall effect size, $\epsilon_i$ is the within-study random error and $\xi_i$ is a random variable reflecting study-specific deviation from the common overall effect. In the literature, it is often assumed that $\epsilon_i$ and $\xi_i$ are independent Gaussian variables, with  $\epsilon_i \sim {\cal N}(0, \sigma_i^2)$ and  $\xi_i \sim {\cal N}(0, \tau^2)$,  where $\sigma_i^2$ is the within-study error variance and depends on the sample size $n_i$ of study $i$. For instance, $\sigma_i^2$ is usually in the order of $n_i^{-1}$. The between-study variance $\tau^2$ reflects  the heterogeneity in the effect sizes across studies. Under the conventional random-effects meta-analysis model, the marginal distribution of $Y_i$ is a Gaussian distribution with mean $\theta$ and variance $\sigma_i^2 + \tau^2$.  The overall effect $\theta$ can be estimated by the  maximum likelihood method \citep{BrockwellandGordon2001, Normand2012}; and the between-study variance $\tau^2$ can also be estimated by the (profile) maximum likelihood method \citep{BrockwellandGordon2001, HardyandThompson1996,Partlett&Riley2017,Langan2019}, DerSimonian and Laird's method of moments \citep{DerSimonianandLaird1986} or quantile approximation \citep{Brockwell2007}.  Equation (\ref{eq:rma}) becomes a fixed-effect meta-analysis model when $\xi_i=0$ ($\mathbb E\,\xi_i=0$ and $\tau^2=0$). 
Within the existing literature, both models have leaned upon a Gaussian assumption. Nonetheless, practical reality often deviates from this parametric assumption, as demonstrated by \citep{Lee&Thompson2008,BrockwellandGordon2001,Sutton&Higgins2008}.

One important objective in meta-analysis is to estimate a confidence interval for the overall effect size. The confidence interval can be  constructed using the asymptotic distribution for the estimator $\hat\theta$ , derived from a maximum likelihood approach  \citep{BrockwellandGordon2001} or a profile likelihood approach \citep{BrockwellandGordon2001, HardyandThompson1996}. Both approaches relied on the assumption that the estimator $\hat\theta$ approximately follows a Gaussian distribution, which however may not be true if  $n_i$ and/or $K$ is small.  Constructing a confidence interval for $\theta$ based on alternative parametric distributions was proposed, including  $t$-distribution, other parametric distributions which allow for potentially skewed and heavy tails in random-effects distributions \citep{Sidik.2002, Lee&Thompson2008, Baker&Jackson2008, Ozturk.2020}, and quantile approximation \citep{Brockwell2007, Jackson&Bowden2009}. Confidence distributions (CDs), which can be viewed as ``distribution estimators" \citep{SchwederandHjort2002}, is a technique which can be used to construct point estimators, confidence intervals and $p$-values. The fusion of CD functions from individual studies have been explored for constructing a confidence interval for the overall effect size \citep{singh2005combining,xie2011confidence}.  Nevertheless, the CD approach still relies on Gaussian assumptions  for the integration of CD functions from individual studies into a unified CD function. For example, \cite{singh2005combining} and \cite{xie2011confidence} used Gaussian distribution to construct CD functions and \cite{Nagashima.2019} proposed a parametric bootstrap algorithm based on Gaussian and $t$-distributions.
We refer the reader to \cite{Veroniki2012} for a  review of methods (all depend on parametric assumptions) to calculate a confidence interval for the estimated overall effect size from a random-effects meta-analysis. 

Different from the existing meta-analysis methods, we consider a more challenging situation where only the $1-\alpha$ confidence interval for the effect size of study $i$, $[L_i,U_i]$, is observed, and individual study effect size estimates $Y_i$ and within-study error variance estimate $\hat \sigma_i^2$ are not observed. Further, we seek to relax the Gaussian assumption.  The aim of this paper is to propose a novel non-parametric meta-analysis method based on empirical likelihood (EL) to make robust inferences on the confidence interval  of the overall effect size $\theta$. We relax the Gaussian distributional assumption for $Y_i$ and $\theta_i$.  In other words, we consider
\begin{equation}
Y_i = \theta_i+ \epsilon_i, \;\;\epsilon_i \sim F_i(\cdot); \qquad
\theta_i = \theta + \xi_i, \;\; \xi_i \sim G(\cdot),\label{eq:rma simulation 2}
\end{equation}
where the distributions $F_i, G$ are not subject to any parametric assumption.
The proposed non-parametric method allows $K \rightarrow \infty$, which is more reasonable than many existing literature where $K$ was treated as a constant. This innovation empowers the non-parametric meta-analysis to effectively manage a substantial number of studies.

The rest of the paper is organised as follows. The proposed non-parametric meta-analysis method is presented in Section \ref{sec:modelbased} and \ref{sec:methods}. We evaluate the performance of the proposed method using simulation studies in Section \ref{sec:sim}. Two real example applications are presented in Section \ref{sec:app}. The paper concludes with a discussion and theoretical proofs for the large sample theory are provided in Web Appendix A of the Supplementary Material.

\section{Model-based meta-analysis approach} \label{sec:modelbased}

For the parametric Gaussian models presented in Section \ref{sec:intro}, \cite{singh2005combining} and \cite{xie2011confidence} developed a general framework for combining results from individual studies in a meta-analysis. They used a Confidence Distribution (CD) function $H(\boldsymbol X, \theta)$ of the data $\boldsymbol X$ and parameter $\theta$, from which inference on confidence intervals, point estimates or $p$-values can be derived. Their method amalgamates the individual study's CD functions $H_i(\boldsymbol X_i, \theta)$ into a unified CD function $H_c(\theta)$, through a monotonic function $g_c(H_1(\boldsymbol X_1, \theta), \cdots, H_K(\boldsymbol X_K, \theta))$. Under the Gaussian assumption in the random-effects meta-analysis model, and assuming the estimated effect size $Y_i$ and the within-study error variance $ \sigma_i^2$ for studiy $i$ are available, a typical CD function for $\theta$ is 
\begin{equation}\label{eq:largesamplesingle}
     H_i(\theta) = \Phi\left(\frac{\theta-Y_i}{ s_i}\right),
\end{equation}
where $\Phi(\cdot)$ is the cumulative Gaussian distribution function, $s_i^2 = {\sigma}_i^2+\hat \tau^2$ in the random-effects model ($s_i^2={\sigma}_i^2$ in the fixed-effect model), and $\hat \tau^2$ is the estimated between-study variance. 

\cite{singh2005combining} and \cite{xie2011confidence} summarized methodologies used in the literature in group decision analysis, where linear combination of individual study result is commonly used as the final outcome. They introduced the following  linear combination function
\begin{align}\label{eq:def g_c}
g_c(u_1,\cdots, u_K) = \sum_{k=1}^K w_k \,\Phi^{-1}(u_k),
\end{align}
where the weight $w_k = 1/s_k$. Then the combination of CDs for $\theta$ is defined as
\begin{align*}
    H_{c}(\theta) & = \Phi\left(\frac{g_c(H_1(\theta),\cdots, H_K(\theta))}{\sqrt{\sum_{i=1}^K w_i^2}}\right) = \Phi\left(\frac{\sum_{k=1}^K w_k \,\frac{\theta - Y_k}{s_k}}{\sqrt{\sum_{i=1}^K w_i^2}}\right).
\end{align*}
Inference on $\theta$ can then be obtained using the combined CD function $H_c(\theta)$. For instance, the $(1-\beta)$ confidence interval for $\theta$ can be constructed as 
\begin{equation}
    [L_{C},\,U_{C}] = [H_{c}^{-1}(\beta/2),\,H_{c}^{-1}(1-\beta/2)]. 
\end{equation}

\begin{figure}[ht]
    \centering
     \includegraphics[width=0.7\textwidth]{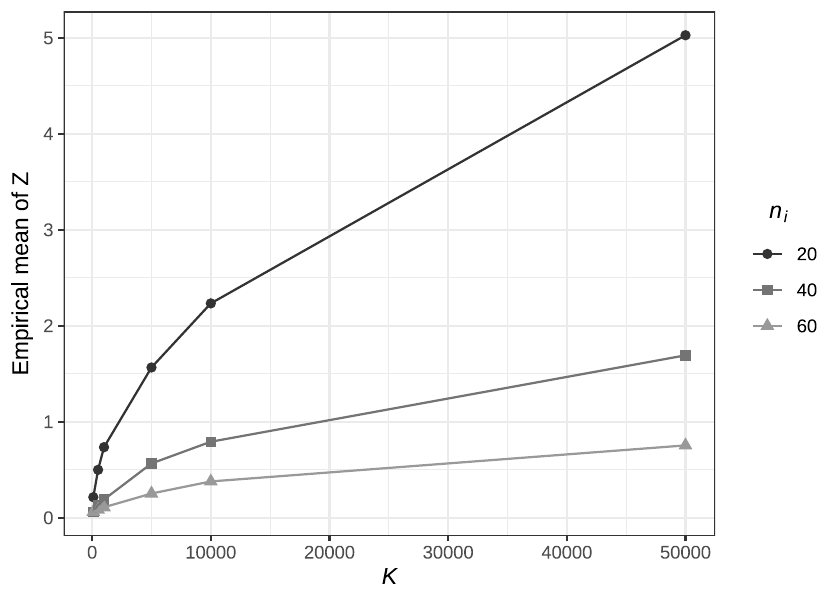}
    \caption{The empirical mean of $Z$ with $\theta_i \sim N(0,1),\, y_{ij}|\theta_i \sim \chi^2(4)-4+\theta_i$ and $Y_i = n_i^{-1}\sum_{j=1}^{n_i}y_{ij}$. We consider the setting with large $K=100, 500, 1000, 5000, 10000, 50000$ and small $n_i=20, 40, 60$ for all $i$.}
    \label{fig:divergence}
\end{figure}

The above approach involves two assumptions:
1. The model error terms in model (\ref{eq:rma}) in Section \ref{sec:intro} follow a Gaussian distribution; 2. The linear combination in (\ref{eq:def g_c}) follows a zero-mean Gaussian distribution. 
If each of the sample size $n_i$ is large enough, the model error term in model (\ref{eq:rma}) can be assumed to be Gaussian, i.e.\ the above assumption 1 is reasonable, but it fails if $n_i$ is small. However, this is a strong assumption. If the first assumption is violated, the second assumption will also be false. If the Gaussian assumption is not held true for some individual studies, in the presence of a large number of studies, i.e.\ large $K$ , the accumulated errors by combining all different studies might not be bounded. 
For instance, under model (\ref{eq:rma simulation 2}) without the Gaussian assumptions, if we consider $\theta_i \sim N(\theta = 0,1)$ but $y_{ij}|\theta_i \sim \chi^2(4)-4+\theta_i, Y_i = n_i^{-1}\sum_{j=1}^{n_i}y_{ij}$, then we will not be able to use the combined statistic $Z=K^{-\frac{1}{2}}\sum_{i=1}^{K}(\theta -Y_i)/s_i$ to construct the meta-analysis confidence interval, since $Z$ will not converge to $0$ when $K\rightarrow \infty$. This can be seen from Figure~\ref{fig:divergence}.  Furthermore, the relationship between the number of studies  $K$ and sample size $n_i$ in meta-analysis has not been extensively discussed. We will propose a non-parametric method and relax this Gaussian assumption in the following section and discuss the relationship between $K$ and $n_i$.

\section{Non-parametric meta-analysis based on empirical likelihood}\label{sec:methods}

In this section, we propose a robust inference procedure for conducting meta-analysis that does not make distributional assumptions on effect size estimates for individual studies. Hence, this new approach is more reliable even if some studies have a small sample size $n_i$.  We first present a generalized theorem and necessary assumptions.

Let $\boldsymbol W_i(\theta)$ be an $r$-dimensional function of $\theta$ for study $i$. If $\boldsymbol W_i(\theta)$ is asymptotically unbiased to $0$ at the true parameter value $\theta_0$,  we can construct the estimating equation $\frac{1}{K}\sum_{i=1}^K \boldsymbol W_i(\theta)=0$ for the unknown parameter $\theta$. 
Following standard conventions in the literature, individual studies are independent of each other. Hence, we assume that the following assumptions hold.
\begin{Assumption}\label{asp:unb}
    \textbf{Asymptotic Unbiaseness}.  For any $i$, at the true parameter $\theta_0$, 
    $$\E\, \boldsymbol W_i(\theta_0)=O(n_i^{-1/2}),$$ which means that each component of $\E\, \boldsymbol W_i$ is in the order of $n_i^{-1/2}$.
\end{Assumption}
\begin{Assumption}\label{asp1}
    \textbf{Independence}. $\boldsymbol W_i(\theta), i=1,\cdots, K$ are independent. 
\end{Assumption}

We then consider the Empirical Likelihood (EL) approach and define the EL function
\begin{equation*}
\mathcal L(\theta)=\sup\left\{\prod_{i=1}^K p_i:\,\sum_{i=1}^K p_i\,\boldsymbol W_i(\theta)=0,\,\sum_{i=1}^K p_i =1\right\}.
\end{equation*}
At the true parameter $\theta_0$, the corresponding EL ratio statistic is
\begin{eqnarray}\label{eq:R}
\mathcal R(\theta_0) = \frac{\mathcal L(\theta_0)}{\sup_{\theta}\mathcal L(\theta)}.
\end{eqnarray}
We further need the following assumptions to discuss the large-sample properties of the EL ratio statistic.
\begin{Assumption}\label{asp:var}
    \textbf{Finite Variance}.  For any $i$, at the true parameter $\theta_0$, $\boldsymbol W_i(\theta_0)$ has a full rank finite covariance matrix $\textbf{M}_i=\text{var } (\boldsymbol W_i(\theta_0))$, and $\textbf{M}_i,i=1,\cdots, K$ are uniformly bounded by $\textbf{M}$, which means for any vector $\textbf{e}\in \mathbb{R}^r$, $\sup_i\textbf{e}^{\T}\,\textbf{M}_i\,\textbf{e}\leq \textbf{e}^{\T}\,\textbf{M}\, \textbf{e}$.
\end{Assumption}

Note that Assumption \ref{asp:var} always holds in practice, since typical classical statistical estimation can always be formalised as a sequence of asymptotically unbiased estimating equations, such as the likelihood inference and generalised estimating equation approaches. The finite variance of $\textbf{M}_i$ is guaranteed when we use standardised estimating equations.

The last assumption is about the relationship between the number of studies $K$ and the individual study sample size $n_i$, which is necessary when we discuss the large sample theories of meta-analysis under a non-parametric approach. To illustrate this assumption, we introduce the following notation and assumption. 
Let $n^{(K)}$ be the minimum number of individual sample sizes, that is $n^{(K)}=\min \{n_i, i=1,\cdots, K\}$. 
\begin{Assumption}\label{asp:rate}
    \textbf{Large Sample Order Condition.} 
    $K/n^{(K)} \rightarrow 0$, as $K\rightarrow \infty$ or $n^{(K)} \rightarrow \infty$.
\end{Assumption}

 Assumption \ref{asp:rate} includes the scenario in the literature, where $K$ is treated as a fixed number and sample sizes $n_i$ may goes to infinity. First, the above assumption allows $K$ to be a fixed number and in this case, as all $n_i$ goes to infinity, the above assumption is always true. In addition, the above assumption is more general since it allows allows $K$, the number of studies, to go to infinity. Assumption \ref{asp:rate} implies that for large values of $K$, we only need $K$ to be in a lower order than all $n_i$, the sample sizes of individual studies. 
 It gives a sufficient condition required for that meta-analyses can be conducted with a substantial multitude of studies. 
Recently, \cite{cai2023individualized} discussed the case of large $K$ but finite $n_i$.

Under these assumptions,  we have the main result as presented in the following theorem, with the proofs given in Web Appendix A of the Supplementary Material.

\begin{theorem}\label{thm:main}
Assume that $\mathcal  R(\theta)$  is an EL ratio statistic 
defined in (\ref{eq:R}). Under Assumptions \ref{asp:unb}-\ref{asp:rate}, as $K\rightarrow\infty$, we have
$
-2\log \mathcal R(\theta_0)\rightarrow\chi^2(r)$ in distribution.
\end{theorem}

For a fixed confidence level $\beta$, we can construct the $(1-\beta)$-level confidence interval for $\theta$ as
\begin{eqnarray}\label{eq:finalCI}
\{\theta : -2\log\mathcal R(\theta)\le \chi^2_{1-\beta}(r)\}.
\end{eqnarray}
Next, we will apply this theorem to two different meta-analysis models. The key point in applying the EL inference is to construct an estimating equation to define the EL ratio statistic.

\subsection{Random-effects meta-analysis}

Consider the random-effects meta-analysis model (\ref{eq:rma simulation 2}).   
Assume that we have collected $1-\alpha$ confidence intervals $[L_i,\, U_i]$ for the parameter $\theta_i$ from $K$ studies. The sample size for each study is denoted as $n_i$, $i=1,\,2,\,\cdots,\,K$. 
Based on the confidence intervals $[L_i,\, U_i]$, we define
\begin{equation}
    W_i(\theta)=\frac{U_i+L_i-2\theta}{U_i-L_i},
\end{equation}
which can be written as $ W_i(\theta)=\frac{\overline{G}_i(\theta)-\underline{G}_i(\theta)}{\overline{G}_i(\theta)+\underline{G}_i(\theta)}$, with $\overline{G}_i(\theta)=U_i-\theta$ and $\underline{G}_i(\theta)=\theta-L_i$.
Then 
\begin{eqnarray*}
\E (\underline{G}_i(\theta_0))
&=&\E\,(\theta_0-Y_i+Y_i-L_i)=\E\,(\E(\theta_0-Y_i)|\theta_i)+\E(Y_i-L_i)\\
&=&\E(\theta_0-\theta_i)+\E(Y_i-L_i)=\E(Y_i-L_i),
\end{eqnarray*}
and $\E (\overline{G}_i(\theta_0))=\E(U_i-Y_i)$. Because of the large sample property of $Y_i$, it is generally true that
\begin{eqnarray*}
\E(Y_i-L_i)\,=\,O(n_i^{-1/2}),\qquad \E(U_i-Y_i)\,=\,O(n_i^{-1/2}).
\end{eqnarray*}
On the other hand, when using empirical likelihood method to construct confidence intervals, we have $\E(Y_i-L_i)-\E(U_i-Y_i)\,\leq c \cdot n_i^{-1}$, for some constant $c$. This holds as well when confidence intervals are constructed via symmetric distributions, such as t-distribution or Gaussian distribution, since $\E(Y_i-L_i)-\E(U_i-Y_i) = 0$. Therefore, we here focus on the confidence intervals such that,
\begin{eqnarray*}
\E(Y_i-L_i)-\E(U_i-Y_i)\leq c \cdot n_i^{-1}.
\end{eqnarray*}
Now we have
\begin{equation}
    \E (\overline{G}_i(\theta_0))-\E (\underline{G}_i(\theta_0))\,=\,O(n_i^{-1}),\qquad \E (\overline{G}_i(\theta_0))+\E (\underline{G}_i(\theta_0))\,=\,O(n_i^{-1/2}),
\end{equation}
and thus Assumption \ref{asp:unb} is satisfied, i.e.,
\begin{equation}
    \E\,W_i(\theta_0)=O(n_i^{-1/2}).
\end{equation}

Now we define
\begin{align*}
\mathcal L(\theta)&=\sup\left\{\prod_{i=1}^K p_i:\,\sum_{i=1}^K p_i\left(\frac{L_i+U_i-2\theta}{U_i-L_i}\right)=0,\,\sum_{i=1}^K p_i =1\right\} , 
\end{align*}
then the corresponding EL ratio statistic is
$
\mathcal R(\theta_0) =  \mathcal L(\theta_0)/\sup_{\theta}\mathcal L(\theta).
$
According to Theorem \ref{thm:main}, we have
$
-2\log \mathcal R(\theta_0)\rightarrow\chi^2(1)$ in distribution.

\subsection{Fixed-effect meta-analysis}
  
We now consider a fixed-effect model, i.e.\ for all $i=1, \ldots, K$, $\theta_i=\theta$ in equation (\ref{eq:rma simulation 2}). Since this is a special case of the random-effects model, we can still use the method presented in the previous subsection to make inferences. In the fixed-effect model, we can also use the confidence level to construct the estimating equation. First, we define 
\begin{eqnarray}\label{eq:est_alpha}
    V_i(\theta)=I\{L_i\leq \theta\leq U_i\}-(1-\alpha).
\end{eqnarray}
Most of the time, the confidence interval $[L_i,\, U_i]$ is constructed based on an asymptotic distribution rather than an accurate distribution, hence, the expectation of (\ref{eq:est_alpha}) is approximately equal to $0$. For example, when the confidence interval $[L_i, U_i]$ was constructed according to the central limit theorem, we have 
\begin{eqnarray*}
\E\,V_i(\theta_0)=\E\,(I\{L_i\leq \theta_0\leq U_i\}-(1-\alpha))=O(n_i^{-1/2}).
\end{eqnarray*}
However, in the random-effects meta-analysis model the above $V_i(\theta_0)$ is not asymptotically unbiased to $0$. This is demonstrated by the following result under model (\ref{eq:rma}) with the Gaussian assumption for $\epsilon_i$ and $\xi_i$,  
\begin{align*}    
\E\,V_i(\theta_0) 
&=\E\,\left(I\left\{Y_i-z_{1-\frac{\alpha}{2}}\cdot\hat{\sigma}_i\leq \theta_0\leq Y_i+z_{1-\frac{\alpha}{2}}\cdot\hat{\sigma}_i\right\}\right) -(1-\alpha)\\
&=\E\,\left(\Phi\left(z_{1-\frac{\alpha}{2}}-\frac{\theta_i-\theta_0}{\hat{\sigma}_i}\right)-\Phi\left(-z_{1-\frac{\alpha}{2}}-\frac{\theta_i-\theta_0}{\hat{\sigma}_i}\right)\right)-(1-\alpha).
\end{align*}
This expectation does not converge to $0$, and it cannot be used to construct the asymptotically unbiased estimating equation in the random-effects model. Therefore, we consider the following three ways to define EL ratio statistic in the fixed-effect meta-analysis model. Denote $\boldsymbol V_i(\theta)$ as the  estimating equation for study $i$, and $r$ as the dimension of $\boldsymbol V_i(\theta)$. 
\begin{itemize}
\item Case I: the estimating equation based on the confidence level, i.e.\ the following function
$    V_i(\theta)=I\{L_i\leq \theta\leq U_i\}-(1-\alpha)$.
\item Case II: the estimating equation based on symmetry (this is the same as the random-effects model), i.e.\ using the following function
\begin{equation*}
V_i(\theta)=W_i(\theta)=z_{1-\frac{\alpha}{2}}\,\frac{L_i+U_i-2\theta}{U_i-L_i}.
\end{equation*}
\item Case III: the estimating equation based on confidence level and symmetry.
In this case, $r=2$, and $V_i(\theta)=(V_{i1}(\theta),\,\,V_{i2}(\theta))^{\T}$, where
\begin{equation*}
V_{i1}(\theta)=I\{L_i\leq \theta\leq U_i\}-(1-\alpha),\quad V_{i2}(\theta)=z_{1-\frac{\alpha}{2}}\,\frac{L_i+U_i-2\theta}{U_i-L_i}.
\end{equation*}
\end{itemize}
By choosing an appropriate $V_i(\theta)$, we can define three different empirical likelihood statistics, which allow us to make inferences on the confidence interval in meta-analysis,
\begin{align}\label{eq:different EL}
\mathcal L_{f1}(\theta)&=\sup\left\{\prod_{i=1}^K p_i:\,\sum_{i=1}^K p_iI\{L_i\le \theta \le U_i\}=1-\alpha,\,\sum_{i=1}^K p_i =1\right\}, \nonumber \\
\mathcal L_{f2}(\theta)&=\sup\left\{\prod_{i=1}^K p_i:\,\sum_{i=1}^K p_i\left(\frac{L_i+U_i-2\theta}{U_i-L_i}\right)=0,\,\sum_{i=1}^K p_i =1\right\} , \\
\mathcal L_{f3}(\theta)&= 
       \sup\left\{\prod_{i=1}^K p_i:\,\sum_{i=1}^K p_iI\{L_i\le \theta \le U_i\}=1-\alpha,\,\sum_{i=1}^K p_i\left(\frac{L_i+U_i-2\theta}{U_i-L_i}\right)=0,\,\sum_{i=1}^K p_i =1\right\}. \nonumber 
\end{align}
Then the corresponding empirical likelihood ratio statistics are
\begin{eqnarray*}
\mathcal R_{fk}(\theta_0) = \frac{ \mathcal L_{fk}(\theta_0)}{\sup_{\theta}\mathcal L_{fk}(\theta)},\qquad k=1,2,3
\end{eqnarray*}
respectively. Following Theorem \ref{thm:main}, we have
$
-2\log \mathcal R_{f1}(\theta_0)\rightarrow\chi^2(1)$, $
-2\log \mathcal R_{f2}(\theta_0)\rightarrow\chi^2(1) $ and $
-2\log \mathcal R_{f3}(\theta_0)\rightarrow\chi^2(2)$, in distribution.
Then for a fixed confidence level $\beta$, we can construct the $(1-\beta)$-level confidence interval for $\theta$ based on (\ref{eq:finalCI}).

\section{Simulation Studies}\label{sec:sim}

To demonstrate the performance of the proposed non-parametric method, we conduct a comprehensive comparison against two established methods: the existing CD method and the conventional meta-analysis method. To be specific,  DerSimonian \& Laird method is used to estimate $\tau^2$ in CD method, and  ‘rma’ in the ‘metafor’ R package where we use REML to estimate $\tau^2$.

We consider four different scenarios in this section, which corresponds to model (\ref{eq:rma simulation 2}) but have different distributions for the zero-mean error terms, 
\begin{align}\label{eq:rma simulation}
y_{ij} & \sim F(\cdot\mid \theta_i, \sigma^2), \;\;
Y_i 
= n_i^{-1} \sum_{j=1}^{n_i} y_{ij}, \nonumber \\
\theta_i &= \theta + \xi_i, \;\;\; \xi_i \sim G(\cdot\mid\tau^2),
\end{align}
where $\mathbb E (y_{ij}) = \theta_i$, i.e.\ $\theta_i$ is the expected effect size of study $i$, $Y_i$ is the estimated effect size of study $i$, and $\theta$ is the overall effect size. The parameter $\sigma^2$ is the within-study variance. This model setting is the same as (\ref{eq:rma simulation 2}), where var$(Y_i)$ is in the order of $\sigma_i^2         = \sigma^2/n_i$. For all scenarios, we set $\theta = 0$, 
and $\sigma^2 = 1$, and we consider different values of $K$,  $20, 60$ and  $100$. 

We set the four simulation scenarios as follows, with different distributions $F$ and $G$. 
\begin{itemize}
\item Scenario 1 is the Gaussian distribution case, where $F = {\cal N}(\theta_i, \sigma^2)$ and $G = {\cal N}(0, \tau^2)$. This is equivalent to model (\ref{eq:rma}).
\item  Scenario 2 sets $F = {\cal N}(\theta_i, \sigma^2)$ and $G$ to log-normal distribution with mean $0$ and variance $\tau^2$. The individual study effect $\theta_i$ is not Gaussian. This is common in practice.  For example, a study's effect size may have a positive and heavy tail. We choose $\xi_i = c_2 \cdot \text{log-normal}(1,1) + d_2$, where $c_2$ and $d_2$ are constants such that the distribution of $\xi_i$ has mean $0$ and variance as a pre-specified value $\tau^2$.
\item Scenario 3 sets $F$ to be non-Gaussian, but $G$ to be Gaussian ${\cal N}(0, \tau^2)$.  When an individual study has a small sample size, the individual studies' effect size estimate $Y_i$ does not follow a Gaussian distribution. 
In our simulation, we first generate $\theta_i\sim {\cal N}(0, \tau^2),\,i=1,\,2,\cdots,\,K$. Given the sample size $n_i$ and $\theta_i$, draw $y_{ij} \mid \theta_i \sim a_{3,i}\cdot \chi^2(4) + b_{3,i}$ such that $\E (y_{ij}\mid\theta_i) = \theta_i, \text{var}(y_{ij} \mid \theta_i) = \sigma^2$. 
\item Scenario 4 is the most general case where we set both $F$ and $G$ to be non-Gaussian. We choose $G$ to be the distribution of $ c_4 \cdot \chi^2(4) + d_4$ to generate $\theta_i$,
and we obtain random sample $Y_i = n_i^{-1}\sum_{j=1}^{n_i} y_{ij}$ with  $y_{ij} \mid \theta_i \sim a_{4,i}\cdot \text{log-normal}(1,1) + b_{4,i}$, where $a_{4,i}$ and $b_{4,i}$ are such that $\E (y_{ij}\mid\theta_i) = \theta_i,$ and $ \text{var}(y_{ij} \mid \theta_i) = \sigma^2$.
\end{itemize}

For each scenario, we obtain the estimated effect size and the estimated within-study variance by $Y_i = n_i^{-1}\sum_{j=1}^{n_i} y_{ij}$ and $\hat \sigma_i^2 = \sum_{j=1}^{n_i}(y_{ij} - Y_i)^2 / (n_i(n_i-1))$, respectively. Then the $1-\alpha=95\%$ two-side confidence interval $[L_i, U_i]$ of  $\theta$ is generated by either 
\begin{align}\label{eq:simulation CI}
    L_i  = Y_i-z_{1-\alpha/2}\cdot \hat{\sigma}_i,\quad
     U_i  = Y_i+z_{1-\alpha/2}\cdot \hat{\sigma}_i, \qquad n_i\ge 30,
     \end{align}
for large $n_i$, or
\begin{align}\label{eq:simulation CI 2}
    L_i  = Y_i-t_{1-\alpha/2}(n_i-1)\cdot \hat{\sigma}_i,\quad
    U_i  = Y_i+t_{1-\alpha/2}(n_i-1)\cdot \hat{\sigma}_i,  \qquad n_i < 30,
\end{align}
for small $n_i$, where $t_\beta(n)$ is the $\beta$-quantile for the $t$-distribution with degree of freedom $n$.

We consider the settings of different values of $\tau^2$, ranging from $0$ to $100$. When $\tau^2=0$, it becomes a fixed-effect meta-analysis model. Based on the observations $[L_i,\,U_i]$, we compare $1-\beta=95\%$ confidence intervals which are constructed via the conventional meta-analysis method, the CD method and the non-parametric EL method. 

\subsection{Scenarios 1, 2 and 3}
\begin{figure}[htb]
    \centering
    \includegraphics[width=0.9\textwidth]{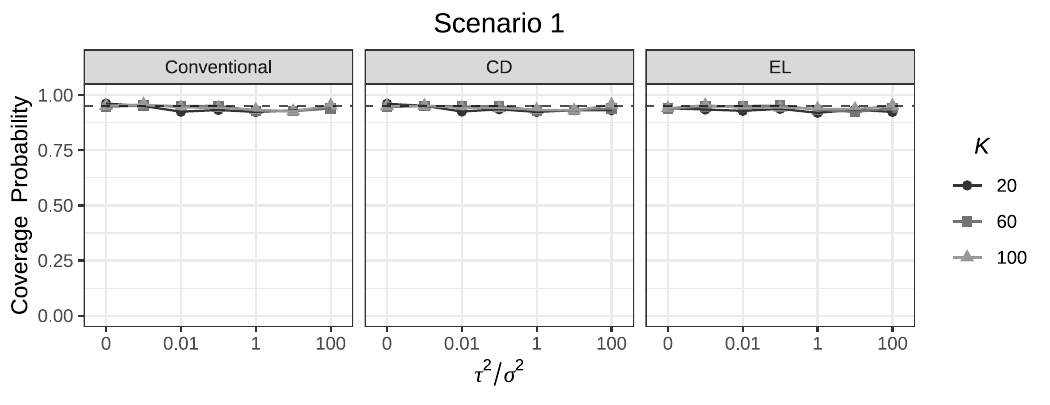}
        \includegraphics[width=0.9\textwidth]{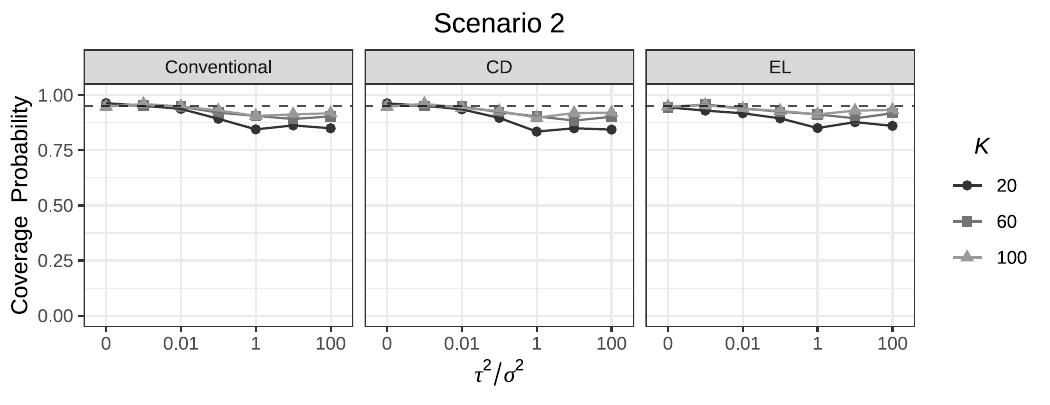}
            \includegraphics[width=0.9\textwidth]{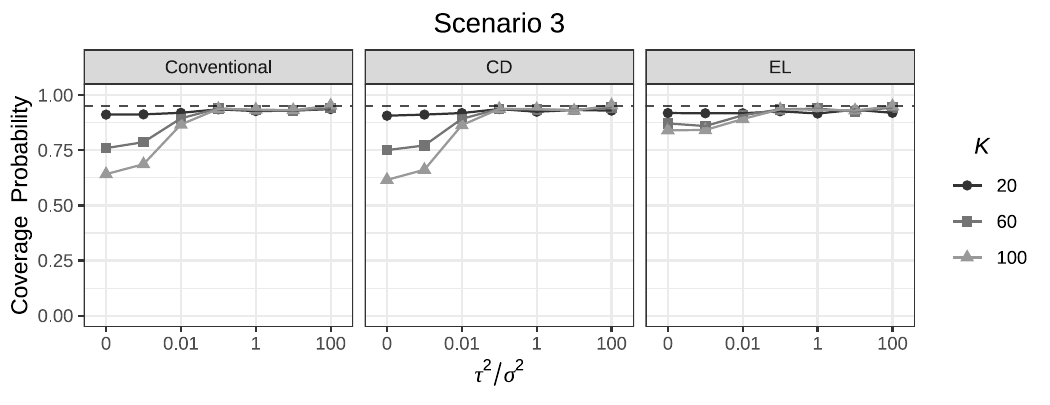}
    \caption{The coverage probabilities of $1-\beta=95\%$ confidence intervals which are constructed via the conventional meta-analysis method, CD method and EL method respectively under the setting of different $K$ and $\tau^2$ in Scenario 1. In each setting the $n_i$ is sampled from $0.2 \cdot {\cal U}[100,500]$.}
    \label{fig:scenario_1}
\end{figure}

We consider the conventional settings of random-effects meta-analysis model $(\tau^2\ne 0)$ and fixed-effect meta-analysis model $(\tau^2 = 0)$. Figure~\ref{fig:scenario_1} shows the results coverage probability against the ratio of between-study variance and within-study variance for Scenarios 1 to 3, based on different values of $K$, 20, 60, and 100, $\sigma^2=1$, and different values of $\tau^2$, $0, 0.001, 0.01, 0.1, 1, 10$ and $100$. For each study, the number of observations $n_i$ is drawn randomly from the uniform distribution $0.2 \cdot {\cal U}[100,500]$. 

The simulation results for Scenario 1 are shown in the first row of plots in Figure \ref{fig:scenario_1}. For all three methods  the coverage probabilities of the confidence interval for the overall effect are close to the nominal level $0.95$, under all different values of $\tau^2$. 
Under the Gaussian assumptions, the non-parametric EL method is as good as the conventional meta-analysis and the CD method.

In Scenario 2, the individual study's effect size estimate is Gaussian but the random-effect is a log-normal distribution. From the second row of plots in Figure \ref{fig:scenario_1}, we can see that all three methods, the conventional meta-analysis, the CD approach and the non-parametric EL method give underestimated coverage probabilities, for large values of $\tau^2$ (with values $ 1, 10, 100$) and small values of $K$.  When $K=20$, the larger $\tau^2$ is, the worse performance for the conventional meta-analysis and the CD method. This is because the Gaussian assumption for the individual study's effect size is invalid and the error is mainly from the non-Gaussian random-effects. When $\tau^2 $ becomes larger, the error of using a Gaussian distribution to approximate a log-normal distribution becomes larger, especially when $K$ is small. The performance of all three methods improves as $K$ increases. For example. when $K=100$, all three methods provide reasonably well coverage probability for overall effect estimate.

The proposed non-parametric EL method exhibits a notable and significant advantage over the other methods in Scenario 3, where the study effect size is a $\chi^2$ distribution but the random-effect is a Gaussian distribution. The results are presented in the third row of plots in Figure  \ref{fig:scenario_1}. The coverage probabilities of confidence intervals derived from both the conventional and CD methods exhibit a rapid decline as the number of studies ($K$) increases, especially when the true $\tau^2$ value is small. This decline implies a notable departure from the anticipated large sample properties for these methods.

In cases where $\tau^2$ is small, it is understood that the primary source of error originates from the within-study approximation. Given that the true distribution for each study in Scenario 3 follows a $\chi^2$ distribution, the within-study error, under the Gaussian assumption, becomes considerably inflated. Consequently, the conventional and CD methods manifest ineffectiveness. However, as the $\tau^2$ value increases, errors in the overall effect estimation stem mainly from the random effects. This scenario proves beneficial for the conventional and CD methods, as the Gaussian approximation aligns more closely with the true distribution of $\theta_i,$ resulting in smaller between-study approximation errors.
In contrast, the non-parametric EL method is more robust and perform better for both small and large values of $\tau^2$. Just as with the conventional and CD methods, there is a potential for convergence challenges within the non-parametric EL method. Notably, as $K$ increases, the convergence probability fails to align with the true nominal value of $0.95$. This deviation can be attributed to the insufficiency of the sample size ($n_i$) within each study, as governed by Assumption \ref{asp:rate}. Further exploration of the effect of augmenting the sample size ($n_i$) will be undertaken in the subsequent subsection, focusing on Scenario 4.

\subsection{Scenario 4 and results based on different $n_i$}
\begin{figure}[htb]
    \centering
        \includegraphics[width=0.9\textwidth]{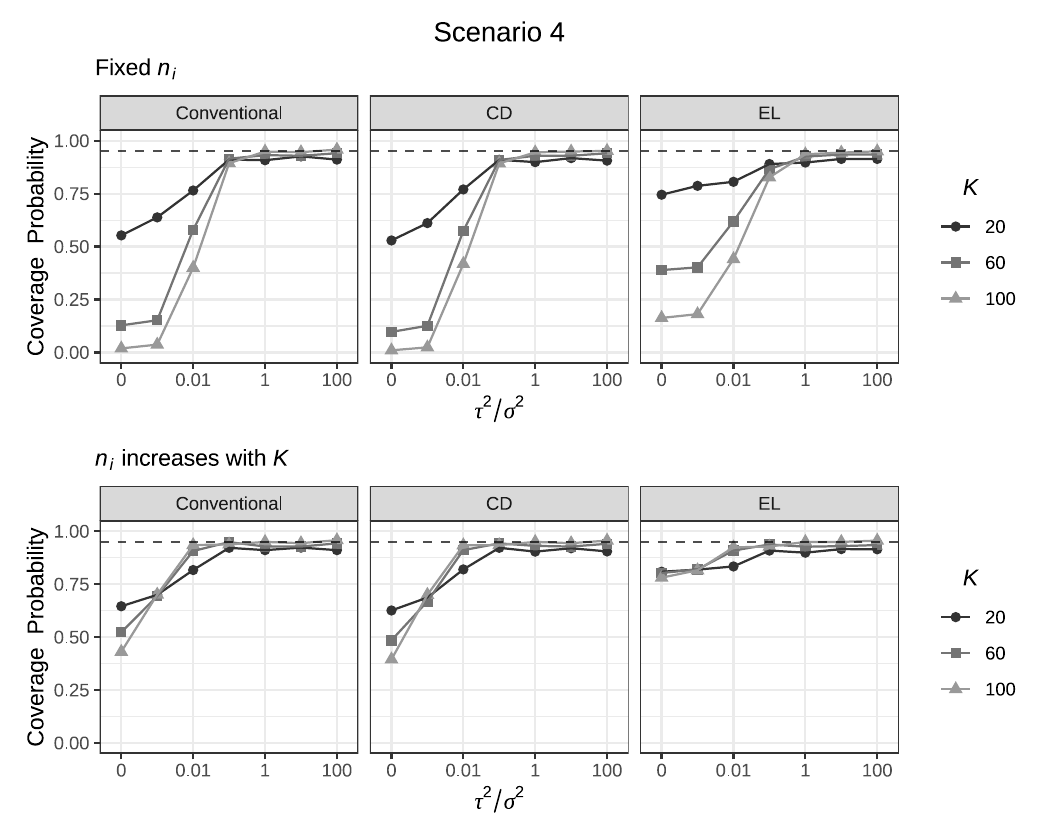}
    \caption{The coverage probabilities of $1-\beta=95\%$ confidence intervals which are constructed via conventional meta-analysis method, CD method and EL method respectively under the setting of different $K$ and $\tau^2$ in Scenario 4. In each setting the $n_i$ is sampled from $0.2 \cdot \mathcal{U}[100, 500]$ (top panel) or sampled from ${\cal U}[4K, 5K]$ (bottom panel).}
    \label{fig:scenario_4}
\end{figure}

We now study the performance of different methods under Scenario 4, that both the individual study effect size and random-effects are non-Gaussian. The top-row plots within Figure \ref{fig:scenario_4} present results for relatively small $n_i$ values, drawn from $0.2 \cdot \mathcal{U}[100, 500].$ It's evident that neither the conventional meta-analysis method nor the CD method performs well under small $\tau^2$ values.  Although the non-parametric EL method exhibits an improved coverage probability,  it still falls considerably short of the target nominal value of $0.95$. This performance gap can be attributed to the relatively modest $n_i$ values presented in this scenario. Significantly larger values of $n_i$ are needed to guarantee the convergence of the log-likelihood ratio statistic. The bottom row of plots in  Figure \ref{fig:scenario_4} shows the performance with large  values of $n_i$, which are random values from ${\cal U}[4K, 5K]$. We can see that the coverage probabilities based on the conventional meta-analysis method and the CD method converge very slowly to the true nominal value $0.95$, even if we have a large number of $n_i$ and a large number of $K$. For instance, when $\tau^2=0, K=100$ and $n_i$ is drawn from ${\cal U}[400, 500]$, both the conventional method and the CD method give coverage probabilities about $0.4$. Remarkably, the non-parametric EL method exhibits significantly improved coverage probabilities, outperforming the conventional and CD methods. Under the same conditions, the non-parametric EL method approximates a coverage probability of about $0.8,$ converging much faster to the true nominal value. 

\subsection{The relationship between $K$ and $n_i$ in the Scenario 4}
\begin{figure}[htb]
    \centering
    \includegraphics[width=0.9\textwidth]{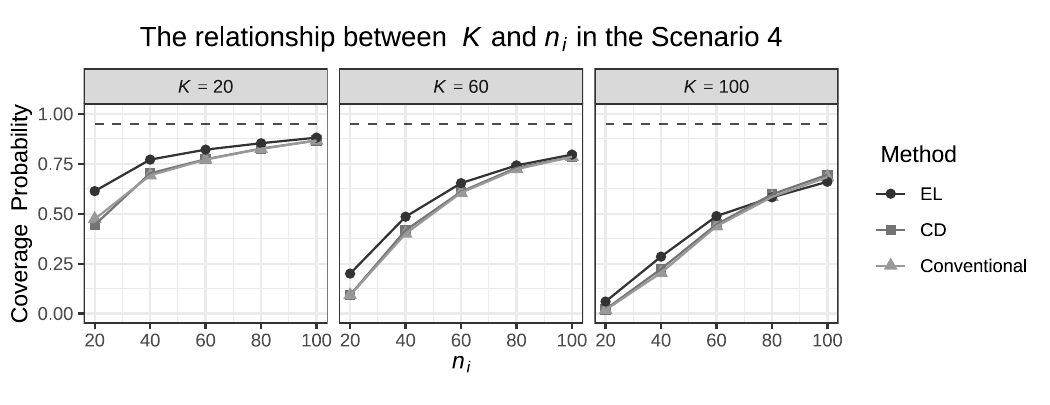}
    \caption{The coverage probabilities of $1-\beta=95\%$ confidence intervals which are constructed via conventional meta-analysis method, CD method and EL method respectively under the setting of different $K$ and $n_i$ in Scenario 4. In each setting the $n_i$ is the same for each $i$ and $\tau^2 = 0.01$.}
    \label{fig:K_and_n}
\end{figure}

Further, we study the relationship between $K$ and $n_i$ in different methods under Scenario 4 with fixed $\tau^2 = 0.01$. The simulation results are show in Figure~\ref{fig:K_and_n}. When $K$ is fixed, the coverage probability of each method will converge to the true nominal value as $n_i$ increases, but our EL method performs better and robustly especially in the small $n_i$. When $n_i$ is fixed, the coverage probability of each method will not converge as $K$ increases since Assumption \ref{asp:rate} is not satisfied.

\subsection{Convergence of the non-parametric EL estimator}
Assumption \ref{asp:rate} was not well addressed in the meta-analysis literature, but it was the key condition to guarantee the consistency of the meta-analysis estimate. Here we further use a numerical study to demonstrate the convergence of the distribution of the log-likelihood function at the true parameter, under different settings of $K$ and $n_i$, i.e.\ numerical justification of Theorem \ref{thm:main}. For brevity, here we only consider the convergence of different methods mentioned in the fixed-effect model. We use QQ plots to empirically examine the accuracy of the $\chi^2(r)$ approximation to the distribution of the log-likelihood function at the true parameter, $-2\log\mathcal{R}(\theta_0)$.

\begin{figure}[htb]
    \centering
     \includegraphics[width=0.8\textwidth]{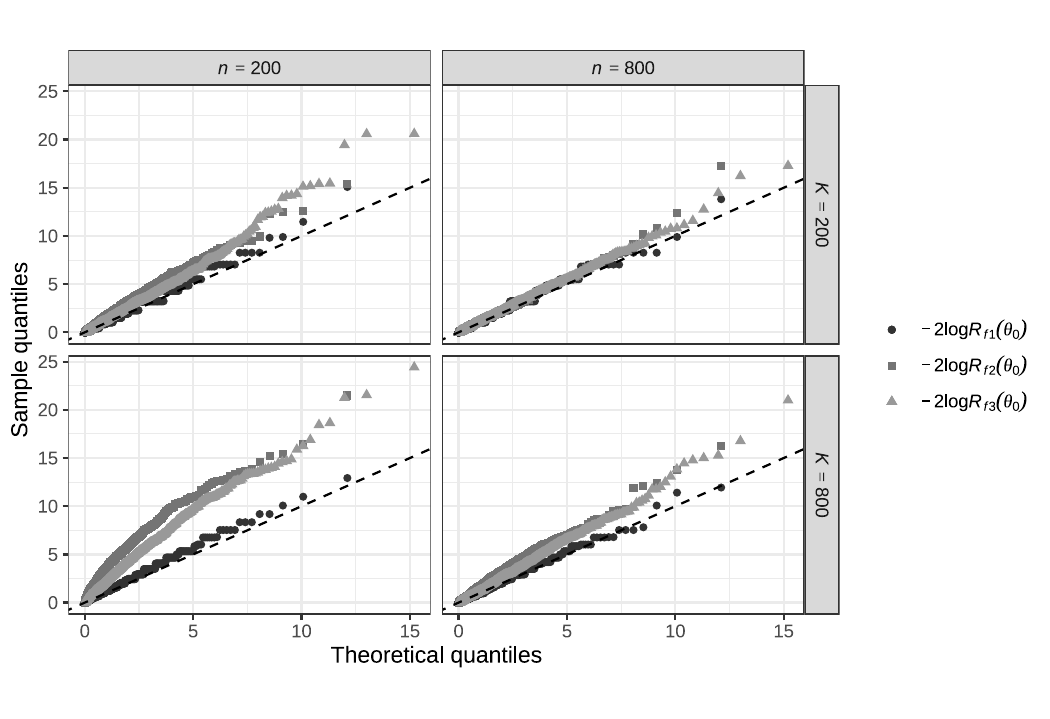}
    \caption{A QQ plot with sample quantiles obtained from $1000$ replicated values of  $-2\log\mathcal{R}(\theta_0)$ under different methods (EL1, EL2 and EL3) and the corresponding the quantiles of the $\chi^2(r)$ distribution ($r=1$ for EL1 and EL2, $r=2$ for EL3) under different settings: $(n, K) = (200,200), (200, 800), (800,200), (800, 800)$. The 45-degree line is included in each plot.}
    \label{fig:qqplot_chisq}
\end{figure}

Taking $F(\cdot\,\mid\,\theta_i)=\chi^2(4),\,\theta_i=\theta_0=4$ for the fixed-effect meta-analysis model, and $n_1=n_2=\cdots=n_K=n$, we compare the sample quantiles obtained from $1000$ replicated values of $-2\log\mathcal{R}(\theta_0)$ under different methods. The theoretical asymptotic distributions of $-2\log\mathcal{R}(\theta_0)$ under EL1, EL2 and EL3 are $\chi^2(r)$ distributions, with $r=1$ for EL1 and EL2, and $r=2$ for EL3. The sample quantiles are plotted in Figure~\ref{fig:qqplot_chisq}. The proximity of sample quantiles to the 45-degree straight line in QQ plots indicates the degree of resemblance between the distribution of these samples and the $\chi^2(r)$ distribution. From these QQ plots we can conclude that, when $K/n = o(1)$, except for extreme upper quantiles, the distributions of $-2\log\mathcal{R}(\theta_0)$ under different methods are reasonably well approximated by a $\chi^2(r)$ distribution. A specific instance can be observed in the top-right panel of Figure \ref{fig:qqplot_chisq}, where $n=800$ and $K=200$. Here, the alignment between sample quantiles and theoretical quantiles is strikingly close.  When $K/n$ increases, especially when $K=800$ and $n=200$, the convergence
of  $-2\log\mathcal{R}(\theta_0)$ to $\chi^2(r)$ diminishes. It is shown by the left-bottom plot of Figure~\ref{fig:qqplot_chisq}, that $-2\log\mathcal{R}(\theta_0)$ does not converge to $\chi^2(r)$, since Assumption \ref{asp:rate} may be violated. When $n=K$ (the top-left plot and the bottom-right plot), $-2\log\mathcal{R}(\theta_0)$ seems still to converge to $\chi^2(r)$, but its performance is not as good as the result with $n=800$ and $K=200$. 

\section{Application} \label{sec:app}

We apply our proposed methods to two real examples. Example 1 has a large number of studies ($K=74$), whilst Example 2 has a small number of studies ($K=6$).  The assumption for $K$ to be bounded in the existing methodologies may not be valid for  a large value $K$ such as in Example 1, since the accumulated errors from each Gaussian approximation within each study may be too large. For a small number of $K$ such as in Example 2, the Gaussian assumption for $\theta_i$ in the conventional meta-analysis methods may not be valid. 

\subsection{Example 1: Mortality in patients with non-COVID illnesses during and before pandemic}

During the COVID-19 pandemic, healthcare providers struggled to manage and deliver quality health care services. Quantifying the effect of COVID-19 pandemic on patients with non-COVID illnesses and on healthcare systems compared to pre-pandemic times can help to understand the impact and disruption of pandemic on health care. A systematic review and meta-analysis of mortality in patients with non-COVID illness during and before the COVID-19 pandemic was conducted \citep{lau2022noncovid}, which presented a meta-analysis (74 observational studies with 491,862 patients), and showed an increased mortality rate during COVID-19 pandemic compared with a pre-pandenmic period for patients with non-COVID illnesses. They studied the  log-risk ratio (RR), where
$$
\text{risk ratio} = \frac{\text{risk  during the pandemic}}{\text{risk  before the pandemic}},
$$
where risk is defined as the ratio of the number of death and the number of patients; and the log-odds ratio (OR), where
$$
\text{odds ratio} = \frac{\text{odds  during the pandemic}}{\text{odds  before the pandemic}},
$$
where the odds is the ratio of the number of death and the number of alive.

Table~\ref{tab:realdata_1} shows the confidence intervals for the estimated overall effect. We also used the the CD function method and our proposed non-parametric empirical likelihood method.  The pandemic is associated with an increased risk of mortality,  with consistent results across the conventional meta-analysis, CD and EL methods as well as fixed-effect and random-effects meta-analysis. Our EL approach resulted in larger CIs ($\mbox{RR:} \; 1.206 \; (1.104, 1.469)$, $\mbox{OR:}\; 1.537, \; (1.293, 2.070)$), which is expected as our method is non-parametric.

Due to the heterogeneity of different studies, a random-effects model is more realistic.  Table~\ref{tab:realdata_1} shows that, when the CD method is used,  the risk ratio is 1.384 (95\%CI: 1.278 to 1.500) from the random-effects model, and is attenuated to 1.075 (95\%CI: 1.068 to 1.082) when we used a fixed-effect model.  However, when using our EL method, the risk ratios are closer to each other between the random-effects and fixed-effect meta-analyses.  In particular, the non-parametric approach with empirical likelihood ${\cal L}_{f2}$ in (\ref{eq:different EL}) can be used in both random-effects model and fixed-effect model. The results are similar despite under different model assumptions.  We observed similar patterns for OR in this example.

\begin{table}[htb]
    \def~{\hphantom{0}}
    \centering
    \caption{Example 1. The combined confidence interval under each method based on the confidence interval of log-risk ratio or log-odds ratio in each study reported in \protect\cite{lau2022noncovid}. Combined confidence intervals were estimated by the conventional meta-analysis, CD method, EL methods in random-effects model and fixed-effect model. The conventional  meta-analysis was fitted by using `rma' in the `metafor' {\bf R}, and for the conventional random-effect meta-analysis we used REML method to estimate the between-study variance $\tau^2$.}
    \label{tab:realdata_1}
    \begin{tabular}{ccccccc}
    \toprule
     & \multicolumn{3}{c}{Fixed-effect meta-analysis} & \multicolumn{3}{c}{Random-effects meta-analysis} \\
     \cmidrule(lr){2-4}\cmidrule(lr){5-7}
     & Method & Est. & 95\% CI & Method & Est. & 95\% CI\\
    \midrule
\multirow{5}{*}{Risk Ratio} & Conventional & 1.075 & $(1.068, 1.082)$ & Conventional  & 1.455  & $(1.270,1.668)$  \\
                            & CD           & 1.075 & $(1.068, 1.082)$ & CD            & 1.384  & $(1.278,1.500)$  \\
                            & EL1          & 1.166 & $(1.086, 1.443)$ & EL            & 1.206  & $(1.104,1.469)$  \\
                            & EL2          & 1.206 & $(1.104, 1.469)$ &               &        &                  \\ 
                            & EL3          & 1.175 & $(1.117, 1.447)$ &               &        &                  \\ 
    \midrule
    \multirow{5}{*}{Odds Ratio} & Conventional & 1.318 & $(1.284, 1.353)$ & Conventional & 1.647 & $(1.407,1.929)$   \\
                                & CD           & 1.318 & $(1.284, 1.353)$ & CD           & 1.650 & $(1.401,1.944)$   \\
                                & EL1          & 1.397 & $(1.385, 1.547)$ & EL           & 1.537 & $(1.293,2.070)$   \\
                                & EL2          & 1.537 & $(1.293, 2.070)$ &              &       &                   \\
                                & EL3          & 1.423 & $(1.385, 1.577)$ &              &       &                   \\
    \bottomrule
    \end{tabular}
\end{table}

\subsection{Example 2: Stillbirths in pregnancy people with and without COVID-19 vaccination}

During the pandemic, COVID-19 vaccination was developed and rolled out at an unprecedented scale and speed addressing the global public health emergency. However,  COVID-19 vaccination hesitancy has been identified as an issue. Apprehension surrounding pregnancy was identified as one contributing source of vaccination hesitancy \citep{Bhattacharya2022, Golder2023, Razain2021}.  Evidence of the effectiveness of COVID-19 vaccination in pregnant women can help to improve the uptake of the vaccination in this vulnerable group. Stillbirth is an important perinatal outcome, which is the death or loss of a baby before or during delivery.  The impact of  COVID-19 mRNA vaccination on stillbirth was evaluated in a systematic review   \citep{prasad2022systematic}  and a meta-analysis of  six studies (66,067 vaccinated v.s.\ 424,624 unvaccinated pregnant people).

Table~\ref{tab:realdata_2} showed that results are inconsistent across methods. From the results obtained from the CD method, COVID-19 vaccination is associated with lower risk of stillbirths. However, in other methods (conventional random-effects meta-analysis, EL random-effects meta-analysis and EL fixed-effect meta-analysis), we found no evidence for the association of COVID-19 vaccination with stillbirths. In summary, the CD method implies that the COVID-19 vaccination is associated with a reduced risk of still-birth. However,  the results from meta-analysis using the non-parametric EL method and the conventional meta-analysis method cannot confirm this. 
\begin{table}[htb]
    \def~{\hphantom{0}}
    \centering
    \caption{Example 2. Stillbirths in pregnancy people with and without COVID-19 vaccination. confidence intervals were estimated by using the conventional random-effects and fixed-effect meta-analysis, the CD method and different non-parametric empirical likelihood methods (EL), based on the confidence interval of log-risk ratio or log-odds ratio in each study reported in \protect\cite{prasad2022systematic}.}
    \label{tab:realdata_2}
    \begin{tabular}{ccccccc}
    \toprule
     & \multicolumn{3}{c}{Fixed-effect meta-analysis} & \multicolumn{3}{c}{Random-effects meta-analysis} \\
     \cmidrule(lr){2-4}\cmidrule(lr){5-7}
     & Method & Est. & 95\% CI & Method & Est. & 95\% CI\\
    \midrule
    \multirow{5}{*}{Risk Ratio} & Conventional & 0.859 & $(0.739, 0.997)$ & Conventional         & 0.859                & $(0.737,1.001)$        \\
                                & CD           & 0.859 & $(0.739, 0.997)$ & CD                   & 0.859                & $(0.739,0.997)$       \\
                                & EL1          & 0.644 & $(0.644, 1.163)$ & EL                   & 0.867                & $(0.762,1.030)$       \\
                                & EL2          & 0.867 & $(0.762, 1.030)$ &                      &                      &                        \\ 
                                & EL3          & 0.854 & $(0.736, 1.080)$ &                      &                      &                        \\ 
    \midrule
    \multirow{5}{*}{Odds Ratio} & Conventional & 0.858 & $(0.738, 0.997)$ & Conventional         & 0.859                & $(0.736,1.001)$        \\
                                & CD           & 0.858 & $(0.738, 0.997)$ & CD                   & 0.858                & $(0.738,0.997)$        \\
                                & EL1          & 0.643 & $(0.643, 1.163)$ & EL                   & 0.867                & $(0.761,1.030)$        \\
                                & EL2          & 0.867 & $(0.761, 1.030)$ &                      &                      &                        \\
                                & EL3          & 0.856 & $(0.735, 1.080)$ &                      &                      &                        \\                         
    \bottomrule
    \end{tabular}
\end{table}

\section{Discussion }
This paper introduces a novel non-parametric method for conducting meta-analysis of confidence intervals. The approach is rooted in empirical likelihood, applicable to both fixed-effect and random-effects meta-analysis models. Notably, the non-parametric EL method departs from the conventional Gaussian assumption often relied upon, while still yielding large sample theoretical outcomes. Our contribution provides the solutions in addressing the requirements regarding the number of studies ($K$) and the sample size ($n_i$) within each study, to ensure the consistency and asymptotic normality of the overall effect estimate. This theoretical result fills a crucial gap that has not been addressed in the current landscape of meta-analysis literature.

The distinguish feature of the non-parametric EL approach is in its reliance on an asymptotically unbiased estimating equation for the unknown parameter, derived from the confidence intervals obtained from individual studies. As long as the function $\boldsymbol W_i(\theta)$ within the estimating equation $\boldsymbol W_i(\theta)=\boldsymbol 0$ remains asymptotically unbiased at a rate $\mathbb E \boldsymbol W_i(\theta_0) = O(n_i^{-1/2})$ (as per Assumption \ref{asp:unb}), the non-parametric EL approach guarantees a consistent estimate. This assumption, being quite mild, holds true for a wide array of existing statistical estimators, such as the maximum likelihood estimator derived from score functions, the estimating equations in Cox partial likelihood for survival analysis, and the generalized estimating equation approach.

In our illustrative examples, our non-parametric EL method produced wider 95\%CI for the overall effect. In example 2, the confidence intervals estimated from the EL method are consistent across methods, in contrast, confidence intervals estimated by CD methods or the conventional meta-analysis are not consistent between the fixed-effect  random-effect models. In particular, it reveals distinctive insights that diverge from existing meta-analysis techniques. In Example 2,  the non-parametric EL method leads to different conclusion -  there is insufficient evidence that COVID-19 vaccination is associated with the risk of stillbirths. This exemplifies the effectiveness of the proposed approach in addressing the underlying challenge in meta-analysis, particularly when individual study estimates  violate the Gaussian assumptions.

\bibliographystyle{biorefs}

\end{document}